\documentclass[%
 reprint,
preprint,
onecolumn,notitlepage,
 amsmath,amssymb,
 aps,
pre,
]{revtex4-2}

\usepackage{enumerate}
\usepackage{multirow}
\usepackage{physics}
\usepackage{graphicx}
\usepackage{dcolumn}
\usepackage{bm}
\usepackage{bbm}
\usepackage{dsfont}
\usepackage{color}
\usepackage[normalem]{ulem}
\usepackage{soul}

\usepackage{cleveref}

\newcommand{\micron}{\mu\mathrm{m}}
\newcommand{\second}{\mathrm{s}}

\begin{document}

\preprint{APS/123-QED}

\title{Dynamic modes of morphogen transport}
\author{Daniel Aguilar-Hidalgo$^{1,2}$}\altaffiliation[Present address: ]{School of Biomedical Engineering, University of British Columbia, Vancouver, British Columbia, Canada V6T 1Z3. Michael Smith Laboratories, University of British Columbia, Vancouver, British Columbia, Canada V6T 1Z4}
\author{Zena Hadjivasilou$^{1,2}$}
\author{Maria Romanova-Michaelides$^{2}$}
\author{Marcos Gonz\'alez-Gait\'an$^{2}$}\email{marcos.gonzalez@unige.ch}
\author{Frank J\"ulicher$^{1}$}\email{julicher@pks.mpg.de}
\affiliation{$^1$Max Planck Institute for the Physics of Complex Systems N\"othnitzer Stra\ss e 38, 01187 Dresden, Germany}
\affiliation{$^2$Department of Biochemistry, Faculty of Sciences, University of Geneva, Geneva, Switzerland}





 \begin{abstract}
Morphogens are secreted signaling molecules that mediate tissue patterning and growth of embryonic tissues.
They are secreted in a localized region and spread through the tissue to form a graded concentration profile. 
We present a cell-based model of morphogen spreading that combines secretion in a local source, 
extracellular diffusion and cellular trafficking. We bring the concept of eigen-modes to the problem of gradient formation to introduce hydrodynamic modes of morphogen transport and
characterize the dynamics of transport by dispersion relations of these dynamic eigenmodes. These dispersion 
relations specify the characteristic
relaxation time of a mode as a function of its wavelength. In a simple model we distinguish two distinct dynamic modes characterized by different timescales. We find that the slower mode defines the effective diffusion and degradation as well as the shape of the concentration profile in steady state. Using our approach we discuss mechanisms 
of morphogen transport in the developing wing imaginal disc of the fruit fly \textit{Drosophila}, distinguishing three 
transport regimes: transport by extracellular diffusion,  transport by transcytosis and a regime where both 
transport mechanisms are combined.   
 \end{abstract}


\maketitle


\section{Introduction}

The development of embryonic tissues implicates the collective organization of a large number of cells in space and time.
A key question is how such tissues can robustly acquire a particular pattern of morphological structures. Biochemical signals, such as morphogens, play an important role to regulate these morphogenetic phenomena during development \cite{Waddington:1940tj,Turing:1952vn,Wolpert:1969wu}. 
Morphogens are secreted in a localized region and spread through the tissue to form 
graded concentration profiles. A system in which morphogen gradients have been extensively
studied is the developing fly wing \cite{Eldar:2003vs,Bollenbach:2005ky,Hornung:2005fs,Kruse:2004di,Bollenbach:2007vf,Kruse:2008kr,Bollenbach:2008hv,Wartlick:2011by,Bosch:2017ds,RomanovaMichaelides:2015hj,AguilarHidalgo:2018jsa}. The developing fly wing is an epithelium, a two-dimensional
single layer of cells.
The morphogen Decapentapledgic (Dpp) is secreted along
a stripe of cells in the center of the wing primordium and exhibits graded concentration profiles at each side of the source \cite{Entchev:2000tg,Teleman:2000vq}. Several mechanisms of transport of Dpp in the tissue have been proposed, including
spreading by extracellular diffusion and transcytosis \cite{Kicheva:2007bha,Zhou:2012ia}.  Transcytosis is defined as a transport regime that
involves the internalization of molecules into the cell and their subsequent recycling to the cell surface
at a different position. 

The dynamics of Dpp in the wing primordium has been studied experimentally using
fluorescently labelled Dpp expressed in the normal source region in the developing wing tissue.
Fluorescently labelled Dpp (GFP-Dpp) forms a concentration profile that is well described by an
exponential with a characteristic decay length that ranges up to 8 cell diameters.
The dynamics of Dpp in the tissue can be revealed by fluorescence correlation spectroscopy (FCS) \cite{Zhou:2012ia}
and by fluorescence recovery after photobleaching (FRAP) \cite{Kicheva:2007bha}. In FCS, a laser beam is parked in the
interface between cells and the temporal correlations of the fluctuating fluorescence signal
are measured to estimate the molecular diffusion coefficient \cite{Zhou:2012ia}. In FRAP, fluorescence of GFP-Dpp is bleached 
in a region of interest adjacent to the source of production. The recovery of fluorescence over time
provides information about the effective diffusion coefficients and degradation rates.

FRAP recovery curves in wild type and in endocytosis defective thermosensitive mutants
of {\it dynamin} suggested that Dpp transport is mediated by endocytic trafficking consistent with
transcytosis \cite{Kicheva:2007bha}. In this case the characteristic length of the Dpp profile depends on the effective diffusion coefficient and degradation rate, which themselves are determined by rates of intracellular trafficking.
The effective diffusion coefficient measured by FRAP is fundamentally different to the molecular diffusion coefficient in the
extracellular space as measured by FCS \cite{chauhan_multiscale_2009,recho_theory_2019}. 

However, the shape of the gradient and the dynamics of the FRAP experiments in wild type can be accounted
for by a regime of transport in which extracellular diffusion is dominating and is fast. In this case an extracellular
gradient would form quickly and the recovery in the FRAP experiment is dominated by the accumulation of
molecules intracellularly \cite{Zhou:2012ia}. While the analysis of endocytosis mutants does not truly support an extracellular 
diffusion regime \cite{Kicheva:2007bha}, with the available assays we cannot currently distinguish between the two regimes
of transport, namely a regime where the gradient shape is 
dominated by extracellular diffusion alone and a regime where intracellular
trafficking contributes significantly to the shape of the gradient. 

The difficulty in distinguish between the two regimes stems largely from the fact that the FRAP
recovery curves can be interpreted in different ways depending on which theoretical model is
considered. This has created controversies and divisions in the field with two different perspective in the interpretation of FRAP data to identify between transport mechanisms. In one case the FRAP dynamics is interpreted as revealing an effective diffusion and
an effective degradation, in the other case the FRAP dynamics corresponds only to the dynamics of
accumulation of intracellular molecules that do not return to the extracellular space.

Here, we develop a theoretical framework to capture both extreme regimes of transport as well as 
combinations of the two. This is achieved by considering at the same time
extracellular diffusion, internalization, recycling and degradation in a model based on discrete cells,
intracellular/extracellular pools and different rates of trafficking between them. In this approach
we introduce the concept of hydrodynamic modes of transport which are eigenmodes of the system that decay
with characteristic relaxation times that depend of the wavelength of the mode. From the mode
structure of the system we can identify the effective diffusion coefficient and degradation rate
that govern the slow dynamics and that are not necessarily the same as the diffusion and degradation measured
by FCS or FRAP. Additionally, we find a relation between the  slow mode of transport and the shape of the steady-state gradient. Using this approach we find that the extreme transport regimes correspond to different values of the
trafficking parameters. This framework will help to design the proper assays to parametrize these rates 
and to determine which actual transport regime underlies gradient formation.

 \begin{figure}[]
\centering
\includegraphics[width=0.7\linewidth]{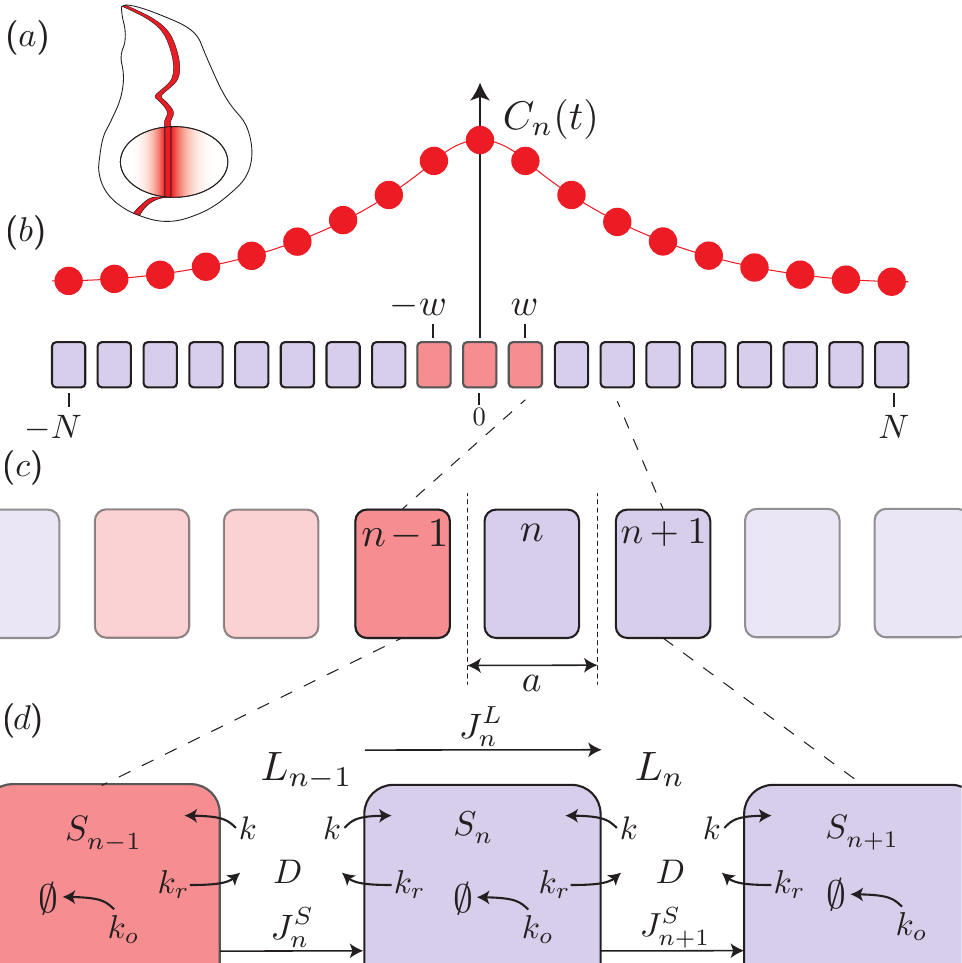}
\caption{\textbf{
Scheme of the morphogen transport model with two compartments.} 
(a) The wing imaginal disk of
the fruit fly is a two-dimensional epithelial sheet with a source
releasing Dpp molecules (red) at the anterior-posterior (AP) compartment boundary (dark red), which is approximately linear. As a result of this localized source and Dpp being degraded in the whole tissue, Dpp shows a concentration gradient in the AP direction, which allows its study in one dimension. 
(b) Discrete profile of morphogen molecules per cell $C_n(t)$, represented by red dots, 
 for an array of $2N+1$ cells (purple and red boxes), from cell number n=-N to cell number n=N. We define a centered morphogen source (red cells) of width $2w+1$. 
(c) We define the cell size $a$ as the distance between two subsequent mid-position of the extracellular space. 
(d) The extracellular pool contains $L_n$ molecules per cell compartment $n$. These molecules can hop directly from one gap between the cells to the adjacent ones at a rate $2D /a^2$. This describes their free diffusion in the extracellular space around the cells with diffusion coefficient $D$, which corresponds to non-directed transport (Brownian motion) in the extracellular space in three dimensions which effectively leads to a diffusion coefficient in one dimension.
Molecules can be internalized with rate $k$ and recycled back with rate $k_r$. The intracellular pool of molecules is denoted $S_n$. Intracellular molecules disappear from the pool at rate $k_o$. Two transport fluxes $J_n^L$ and $J_n^S$ are defined in \cref{teq:fluxnL,teq:fluxnS}.}
\label{tfig:scheme}
\end{figure}

\section{Discrete model for morphogen transport}\label{sec:toy}
We present a general cell-based model for the transport of ligand molecules. The discrete nature of this model allows the analysis of the transport dynamics while preserving intrinsic length-scales of the system such as the cell size, which can be lost in continuum schemes as they represent approximations of the more realistic cell-based model.

\subsection{Dynamic equations of the transport model}
We present a general cell-based model for the transport of ligand molecules that are secreted locally and spread along one axis of the tissue, which consists of a row of cells of size $a$, and specify a morphogen production and secretion of free ligands region of size $(2w+1)a$ placed in the center of the tissue, where ligands enter the extracellular space at flux per cell $\nu$.
We denote by $L_{n}$ the number of ligand molecules in the extracellular space between cell $n$ and cell $n+1$, 
and by $S_n$ the number of molecules in cell $n$, see \cref{tfig:scheme}. 
The dynamic equations for these molecule numbers read:
\begin{align}
\frac{d L_n}{d t}&=\frac{D}{a^2}\left(L_{n+1}-2L_n +L_{n-1}\right)+\frac{k_{r}}{2}\left(S_n+S_{n+1}\right)
-kL_n+\frac{1}{2}\left(\nu_{n}+\nu_{n+1}\right)\label{teq:l}\\
\frac{d S_n}{d t}&=\frac{k}{2}\left(L_{n-1}+L_{n}\right)-k_rS_n-k_oS_n\label{teq:se}\,,
\end{align}
These equations apply for $L_n$ if $-N-1\leq  n \leq N$ and for $S_n$ if $-N\leq  n \leq N$.
Here $D$ is an extracellular diffusion coefficient, $k$ is an internalization rate, $k_{r}$ is a recycling rate, and $k_{o}$ is a degradation rate.

We also need to specify the boundary conditions. When solving the equations for $L_{-N-1}$ and
$L_N$ we use the boundary values 
$S_{-N-1}=S_{-N}$, $S_{N+1}=S_{N}$, together with $L_{-N-2}=L_{-N}$ and $L_{N+1}=L_{N-1}$,
which corresponds to no flux at the boundaries. 

The total number of morphogen molecules per cell is

\begin{equation}\label{teq:totcon}
C_{n}=\frac{1}{2}\left(L_{n-1}+L_{n} \right)+S_{n}\,.
\end{equation}

We will study the system in a finite field of $2N+1$ cells centered around the source of width $(2w+1)a$, which contains $2w+1$ source cells, see \cref{tfig:scheme}. Thus our system constitutes a set of $2(2N+1)$ linear differential equations.

\subsection{Ligand balance due to production, transport and degradation}

We now discuss the balance of ligand molecules due to transport, sources and sinks. The currents of extracellular and intracellular ligands are defined as
\begin{align}\label{teq:fluxnL}
J_n^{L}=&-\frac{D}{a^{2}}\left(L_{n}-L_{n-1}\right)\\
J_n^{S}=&-\frac{k_{r}}{2}\left(S_{n}-S_{n-1}\right)\,.\label{teq:fluxnS}
\end{align}
The dynamic  \cref{teq:l,teq:se} define the balance of molecule number $C_{n}$
\begin{align}\label{teq:balance}
\frac{dC_n}{dt}=&\frac{1}{2}(J_{n-1}^{L}-J_{n+1}^{L})+\frac{1}{2}(J_{n}^{S}-J_{n+1}^{S})
-k_{o}S_{n}+\frac{1}{4}\left(\nu_{n-1}+2\nu_{n}+\nu_{n+1}\right)\,.
\end{align}
This equation confirms the definition of currents $J_n^{L}$ and $J_n^{S}$, and we can identify the degradation rate $k_{o}$ and the effective source term $(\nu_{n-1}+2\nu_{n}+\nu_{n+1})/4$ at cell $n$.

\subsection{Decomposition in hydrodynamic transport modes}\label{sec:soleq}

We use a Fourier representation for the morphogen profiles with no-flux boundary conditions.
The general solution to the dynamic \cref{teq:l,teq:se}  can 
be written as
\begin{equation}\label{eq:sspluspert}
\begin{pmatrix}
L_{n}\\
S_{n}
\end{pmatrix}
=
\begin{pmatrix}
L_{n}^{ss}\\
S_{n}^{ss}
\end{pmatrix}
+
\sum_{\alpha=1}^2
\sum_{m=-N}^N
a_{m}^{\alpha}
\begin{pmatrix}
L_{m}^{\alpha}\\
S_{m}^{\alpha}
\end{pmatrix}
e^{iq_{m}n}
e^{-s_{\alpha}(q_{m})t}\,.
\end{equation}
The boundary conditions used are satisfied for the wave numbers
\begin{equation}\label{eq:Fq}
q_{m}=\frac{2\pi m}{2N+1}\,,
\end{equation}
where $m=-N\ldots N$. 
In \cref{eq:sspluspert}, the time-independent terms are the steady-state profiles $L_{n}^{ss}$ and $S_{n}^{ss}$ which the system
reaches at long times. The time-dependent
terms are relaxation modes of wave number $q_{m}$
and wavelength dependent relaxation rate $s_{\alpha}(q_m)$, where $\alpha=1,2$ 
is a mode index. The mode amplitudes are denoted
$a_{m}^{\alpha}$. They are in general 
complex numbers that obey $a_{-m}^{\alpha}=(a_{m}^{\alpha})^*$,
where the star denotes the complex conjugate. The relaxation rates $s_\alpha$ and the
eigenmodes $(L_{m}^{\alpha},S_{m}^{\alpha})$
follow from an eigenvalue problem:
\begin{equation}\label{teq:eigenprob}
M(iq)
\begin{pmatrix}
L^{\alpha}\\
S^{\alpha}
\end{pmatrix}
=
-s_{\alpha}
\begin{pmatrix}
L^{\alpha}\\
S^{\alpha}
\end{pmatrix}\,.
\end{equation}
Here $M(z)$ with $z=iq$ is the matrix 
\begin{equation}\label{teq:M}
M=\begin{pmatrix}
 -k +\frac{D}{a^{2}}\left(e^{- z}-2+e^{z} \right)&k_{r}
 \left(1+e^{z} \right)/2\\
k \left(1+e^{-z} \right)/2&-k_r-k_o
\end{pmatrix}\,,
\end{equation}
the relaxation rates $s_{\alpha}$ are the eigenvalues of $M$, and $(L^{\alpha},S^{\alpha})$ are the eigenvectors.
The eigenvalue problem (\ref{teq:eigenprob}) is solved using $\text{det}(M-\mathbb{I}s)=0$, where $\mathbb{I}$ is the identity matrix. This equation defines a polynomial in $s$, which is the characteristic polynomial of the eigenvalue problem
\begin{align}\label{teq:poly}
s^{2} -(k + k_{r} + k_{o}) s+k k_{o} -2 \left( \frac{k k_{r}}{4} + \frac{D}{a^2} (k_{r} +  k_{o} - s)\right) (\cosh{z} - 1)=& 0 \quad ,
\end{align}
\Cref{teq:poly} has two solutions for two different values $s_{\alpha}$ of $s$ per wavenumber $z=iq$, which are two 
eigenvalues $s_{\alpha}$ that define two different relaxation times in the transport dynamics at different length-scales. The corresponding eigenvectors $(L^{\alpha},S^{\alpha})$ can then be determined from (\ref{teq:eigenprob}).
The full set of modes $(L_{m}^{\alpha},S_{m}^{\alpha})$ with eigenvalues $s_\alpha(q_m)$
then follows by using $q=q_m$ for all $-N\leq m\leq N$.

\subsection{Steady state concentration profiles}\label{sec:ss}

At long time, the dynamics of the system reaches a  
time-independent steady state.  The steady-state morphogen 
profiles provide the shape of the distribution of molecules in the long-time limit.

For a source with constant production $\nu_{n}=\nu$ for
$-w\leq n \leq w$ and $\nu_{n}=0$ outside the
source region, 
the steady-state solution can be expressed in regions of
constant $\nu$ in the form
\begin{equation}\label{eq:ssww}
\begin{pmatrix}
L_{n}^{ss}\\
S_{n}^{ss}
\end{pmatrix}
=
\begin{pmatrix}
L_{}^{0}\\
S_{}^{0}
\end{pmatrix}
+
\begin{pmatrix}
L_{}^{-}\\
S_{}^{-}
\end{pmatrix}
e^{-n\sigma}
+
\begin{pmatrix}
L_{}^{+}\\
S_{}^{+}
\end{pmatrix}
e^{n\sigma}\,,
\end{equation}
with constant $(L_{}^{0},S_{}^{0})$ in the source region, and where $(L_{}^{\pm},S_{}^{\pm})$ are the amplitude of the positive and negative exponential contributions to the spatial concentration profile, respectively.

We can use this solution to construct the full concentration profile in a piecewise manner. We need to match together the three regions for $-N\leq n\leq -w$ at the left side of the source, $-w\leq n \leq w$ at the source and $ w\leq n\leq N$ at the right side of the source such that they obey the dynamic equations at the source boundaries. 
Additionally, boundary conditions apply as stated above.

The decay rate $\sigma$ is determined from 
the condition $\det(M(z=\sigma))=0$, which holds for steady states. 
We then find
\begin{equation}\label{teq:coshsigma}
\cosh{\sigma}=\frac{1}{2}\left[\frac{k_{r}}{4k_{o}} + \frac{D}{k a^2} \left(1 + \frac{k_{r}}{k_{o}}\right)\right]^{-1}+1 \,.
\end{equation}
We can define the decay length of the graded distribution of molecules outside of the source
\begin{equation}\label{eq:lambdasigma}
\lambda=\frac{a}{\sigma}\,.
\end{equation}

If $\lambda$ is larger than the cell size $a$, the decay length can be approximated as
\begin{equation}\label{teq:lambdaapprox}
\lambda\approx \left[\frac{a^{2}}{4}\frac{k_{r}}{k_{o}} + \frac{D}{k} \left(1 + \frac{k_{r}}{k_{o}}\right) \right]^{1/2}\,.
\end{equation}
We find that the decay length $\lambda$ contains two terms.
The first corresponds to the contribution of recycling at rate $k_r$ to the formation of the gradient in the
absence of diffusion. The second term
describes the effects of extracellular diffusion $D$ and cellular capture with rate $k$ of morphogen molecules by
endocytosis; it also includes effects of recycling, which makes intracellular molecules
available again to diffuse extracellularly.
We will analyze this further in the next section.
See Apendix \ref{Ap:ss} for details.

\subsection{Relaxation time spectrum of the transport equations}

The general solution of the transport equation \cref{eq:sspluspert} expresses the dynamics of the extracellular and intracellular pools of molecules as a superposition of relaxation modes. The corresponding 
relaxation rates are given by $s_{\alpha}(q_m)$, for each wavenumber $q=q_{m}$. 
These relaxation rates read
\begin{widetext}
\begin{align}\label{teq:eigenvals}
s_{1,2}&=\frac{1}{2}(k + k_r + k_o) +\frac{D}{a^{2}}(1 - \cos{q_{}}) \nonumber\\
&\pm \left[\left(\frac{1}{2}(k + k_r + k_o) -\frac{D}{a^{2}}(1 - \cos{q_{}})\right)^2-kk_{o}-2(1 - \cos{q_{}})\left(\frac{kk_{r}}{4}+\frac{D}{a^{2}}(k_{r}+k_{o})  \right)\right]^{1/2}
\end{align}
\end{widetext}
The functions $s_{\alpha}=s_{\alpha}(q_{})$ defined above are the so-called dispersion relations of the propagating system. 
\begin{figure}[]
\centering
\includegraphics[width=1\linewidth]{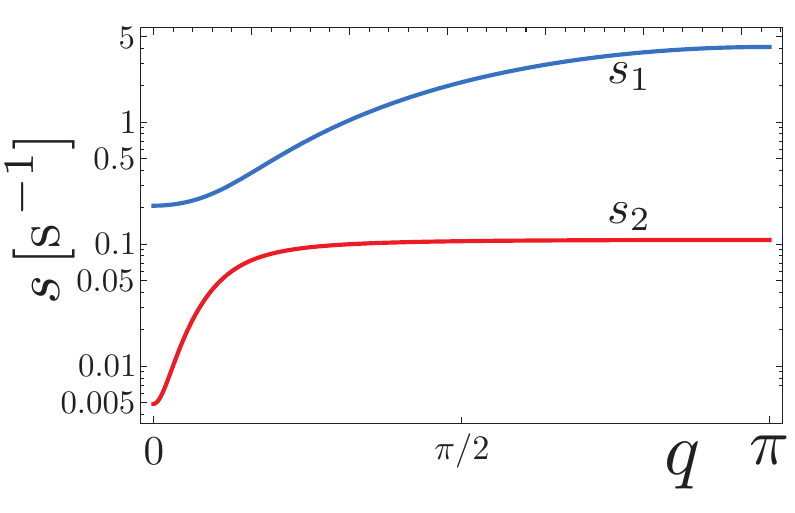}
\caption{\textbf{Example of dispersion relations.}  
Shown are the relaxation rate $s_1$ (blue) of the fast and $s_2$ (red) of the slow relaxation mode
as a function of wave-number $q$. 
Parameter values are $D=1\micron^{2}/\text{s},\,k=0.1/\second,\,k_{r}=0.1/\second,\, k_{o}=0.01/\second,\,a=1\micron$.}
\label{fig:disp_rel_toy}
\end{figure}
 An example of the dispersion relations of the two modes $s_{1,2}$ is shown in \Cref{fig:disp_rel_toy}
 as a function of $q$. 
 The mode $s_{1}$ is faster than $s_{2}$ at all wave-lengths.

\subsection{Effective diffusion constant and effective degradation rate}

The dispersion relations introduced in the previous section carry information about how morphogen profiles 
evolve in time. Of particular interest is the dynamics at long wavelengths (small wavenumber $q_{}$), which provides information about the large scale dynamics of the system. To this end, we expand the 
dispersion relations $s_{\alpha}(q_{})$ as Taylor series in the wavenumber $q$ as
\begin{equation}\label{eq:seriesqsmallq}
s_{\alpha} = K_{\alpha} + \frac{D_{\alpha}}{a^2}q_{}^{2}+ O(q_{}^{4})
\end{equation}
where $K_{\alpha}$ is the effective degradation rate and $D_{\alpha}$ is the effective diffusion coefficient in each dynamic mode.  The expansions \cref{eq:seriesqsmallq} only contain even powers because the transport equations \cref{teq:l,teq:se} do not contain drift terms. 
As a consequence, $q_{}$ appears in the characteristic polynomial \cref{teq:poly} in even functions. 
We assign $\alpha=1$ to the faster mode and $\alpha=2$ to the slower mode.
In \cref{fig:spectrum_toy}, the curvature of $s_{\alpha}(q)$ for small $q_{}$ corresponds to 
the effective diffusion coefficient which can be calculated from \cref{eq:seriesqsmallq} as the 
coefficient of the $q_{}^{2}$-term.
We find that the transport model defined in \cref{teq:l,teq:se} in general exhibits two diffusive modes via which molecules can be transported with effective diffusion coefficients $D_{\alpha}$ and 
effective degradation rates $K_{\alpha}$.

\subsection{Dispersion relations in the complex plane}

\begin{figure}[h!]
\centering
\includegraphics[width=1\linewidth]{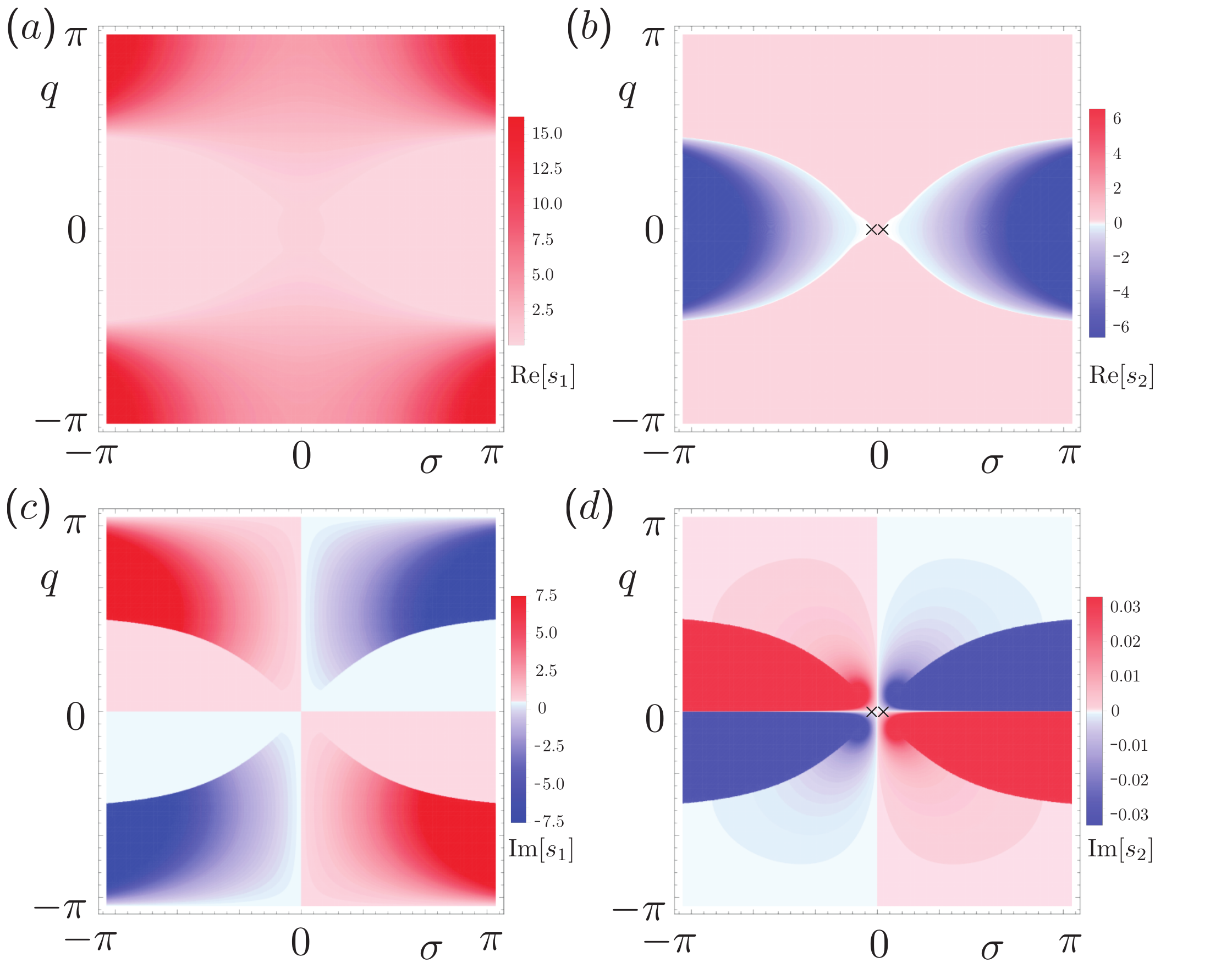}
\caption{\textbf{Dispersion relations in the complex plane} 
(a)-(d) Representation in the complex plane of the real ((a) and (b)) and imaginary ((c) and (d)) parts of the complex relaxation rates $s_{1}$ ((a) and (c)) and $s_{2}$ ((b) and (d)) as a function of complex wave number $z=\sigma+iq$. Values of $z$ for which 
$s_2=0$ are indicated in the real and imaginary parts of $s$ by black crosses.  
Parameters: $D=1\micron^{2}/\second,\,k=0.1/\second,\,k_{r}=0.1/\second,\, k_{o}=0.01/\second,\,a=1\micron$. }
\label{fig:spectrum_toy}
\end{figure}

The dispersion relation not only provides information about the relaxation times for given wave number,
but also provides information about the steady state. The steady state corresponds to infinite relaxation
time $s=0$. In order to find $s=0$ we need to extend wave numbers in the complex plane and write
$z=\sigma+iq_{}$. \Cref{fig:spectrum_toy} shows the real and imaginary parts of $s_{1,2}$ as a function
of complex wave number $z$. 
The figure shows that only the slow mode $s_{2}$ contains a point for which 
\begin{equation}
s_{2}=0\,,
\end{equation}
which corresponds to the steady state. This occurs for real 
$z=\sigma=\pm a/\lambda$, defined by the decay length given in \cref{teq:coshsigma}. Thus, the shape of the distribution of molecules at steady state is determined by the slow relaxation mode $s_{2}$. 
And the dynamics of the gradient formation at long times is governed by the effective diffusion coefficient 
$D_{2}$ and the effective degradation rate $K_{2}$.
The steady state is captured approximately when
expanding  the relaxation rate $s_{2}$ for small $z$ as
\begin{equation}
s_{2}\approx K_{2}-\frac{D_{2}}{a^2}z^2\,.
\end{equation}
The zeros of $s_{2}=0$ correspond to $z=\pm\sigma$ with $\sigma^{2}\approx a^2K_{2}/D_{2}$. This
shows that for decay lengths which are large compared to the cell size we have
\begin{equation}
\lambda\approx\sqrt{\frac{D_{2}}{K_{2}}}\,,
\end{equation}
which corresponds to the decay length for a simple diffusion degradation process.
%

\section{Regimes of morphogen transport}

The transport model introduced above shows two distinct dynamic modes characterized by different timescales. 
We now study limit cases of relevance to discuss experiments considered in the literature on morphogens: 
(i) a scenario of pure transcytosis, in which morphogen transport is driven only via internalization and recycling 
of molecules in the absence of extracellular diffusion, $D=0$, and (ii) a scenario of exclusive extracellular diffusion without transcytosis, $k_{r}=0$.

\subsection{Pure transcytosis: No extracellular diffusion}

In the case without extracellular diffusion, $D=0$, molecules are transported by internalization and recycling.
The dispersion relations are then given by 

\begin{align}\label{teq:eigenvalsD0}
s_{1,2}=&\frac{1}{2}\left(k + k_o + k_r  
\pm \sqrt{\left(k + k_o + k_r \right)^2-4k\left(k_{o}+k_{r}(1 - \cos{q})/2 \right)}\right)\,,
\end{align}
The effective degradation rates are 
\begin{equation}
 K_{1,2}=\frac{1}{2}\left(k + k_o + k_r \pm \sqrt{\left(k + k_o + k_r\right)^2- 4k k_o }\right),\label{teq:Ks}
\end{equation}
and  the effective diffusion coefficients are
\begin{equation}\label{teq:DsD0}
 D_{1,2}=\mp\frac{a^{2}kk_{r}/4}{\sqrt{(k+k_o+k_r)^2-4kk_o}}\,.
\end{equation}
Here $D_2>0$ is the long term effective diffusion coefficient. 
Note that $D_1$ is negative. This does not imply an instability in the discrete model presented here. 
The decay length of the steady-state profile can be approximated as 
\begin{equation}
\lambda\approx \frac{a}{2}\sqrt{\frac{k_{r}}{k_{o}}}\,,
\end{equation}
which depends on the recycling rate $k_r$ and the output rate $k_o$ from the intracellular pool.

\subsection{Pure extracellular diffusion: No recycling of molecules}

In the case where no recycling of molecules occurs $k_r=0$, molecules are transported by extracellular diffusion $D$ only. 
In this case, the dispersion relations are given by 

\begin{align}
s_{1}=&k + 2\frac{D}{a^{2}}(1-\cos{q})\label{teq:disprelkr02} \\
s_{2}=&k_{o} \,,\label{teq:disprelkr0}
\end{align}
where the mode $s_{2}$ is not diffusive. Thus, transport only occurs via extracellular diffusion via the mode $s_1$. 
From \cref{teq:disprelkr0,teq:disprelkr02} we find the effective degradation rates
$K_{1}=k$ and $K_{2}=k_{o}$,
and the effective diffusion coefficients
$D_{1}=D$ and  $D_{2}=0$.

The decay length of the steady-state gradient depends only on the extracellular diffusion coefficient $D$ and on the effective internalization rate $k$, 
\begin{equation}
\lambda\approx\sqrt{\frac{D}{k}}\,.
\end{equation}
Note that contrary to the pure transcytosis case, in the pure extracellular diffusion case $\lambda$ is independent on the cell size.
In this case, molecules internalized to cells do no longer contribute to transport.

\subsection{Extracellular diffusion combined with transcytosis}

In this case, both the extracellular diffusion coefficient and the recycling rates are present. This results in a higher complexity of the transport dynamics. The dispersion relations are given by \cref{teq:eigenvals}.
The effective degradation rates are given by
\begin{equation}
 K_{1,2}=\frac{1}{2}\left(k + k_o + k_r \pm \sqrt{\left(k + k_o + k_r\right)^2- 4k k_o }\right)\label{teq:Ks}
\end{equation}
and the effective diffusion coefficients read
\begin{equation}\label{teq:Ds}
 D_{1,2}=\frac{-a^{2}k k_{r}/4+D(k-K_{2,1})}{K_{1,2}-K_{2,1}} \, .
\end{equation}

The decay length of the steady-state gradient, as commented in section \cref{sec:ss}, contains two independent contributions with dominating recycling rate $k_{r}$ and extracellular diffusion coefficient $D$, see \cref{teq:lambdaapprox}.

We now discuss these three cases within the context of an application of our theoretical framework in comparison to experiments on the fly wing imaginal disc, see \cref{fig:expss}.

\section{Application to experimental data}
Recent literature study the dynamics of fluorescent signals to provide insights on features such as growth control, cellular and molecular patterning, wound repair, and scaling \cite{AguilarHidalgo:2018jsa, Vollmer:2017fr, blasle_quantitative_2018, soh_frap_2018, magny_pegasus_2019, kobb_tension_2017, umulis_organism-scale_2010, almuedo-castillo_scale-invariant_2018}.
The theoretical framework introduced in the sections above can be applied to such studies to find expressions of time- and length-scales from the analysis of the relaxation time spectrum in rationalized coupled linear systems. 
This mode structure describes slow and long wavelength modes in the limit of small amplitudes where the underlying nonlinear system can be linearized.
In the case of morphogen transport, these time- and length-scales read as the effective degradation  rate $K_2$, the effective diffusion coefficient $D_2$ and the decay-length $\lambda$ of the spatial concentration profile. 
These quantities are functions of all the parameters of the transport model, and can be inverted to calculate elementary transport rates ($D,\,k,\,k_r,\,k_o$ in our model) from experimentally determine decay-length and effective dynamics. 
This can be done by tagging the morphogen molecule to a fluorescent molecule and imaging spatial concentration profiles and recording time series of the changes in the fluorescent intensity. 
Our theory describes average behaviors when averaging over many samples and experiments.
It is noteworthy to mention that the transport model can be modified and extended to capture particularities of experimental observations.

\subsection{Application to Dpp in the wing imaginal disc of the fruit fly}

As an example of application of the theory presented in previous section, we introduce a particular case of study and further discussion, which requires the addition of extra components in the model, see Appendix \ref{Ap:comp5} for further example.

\subsubsection{Dynamics of an immobile fraction}

Experimental  data of GFP-Dpp concentration profiles and FRAP studies 
from \cite{Kicheva:2007bha}, see \cref{fig:expss,fig:3regimes}
reveal a so-called immobile fraction \cite{Kicheva:2007bha} which relaxes on long time scales. 
This implies that there is a transfer of molecules from the intracellular mobile pool $S_{n}$ to an immobile 
intracellular pool $S_{n}^{(i)}$ with rate $k_{i}$. Then the total rate of molecules $k_o=k_{i}+k_{1}$ leaving the 
mobile intracellular pool $S_{n}$ is the sum of the immobilization rate $k_{i}$ and the rate at which molecules are degraded from the mobile intracellular pool $k_{1}$. To study the FRAP dynamics we update our transport model with a third equation that captures the dynamics of this immobile intracellular pool $S_{n}^{(i)}$, 
\begin{align}
\frac{d L_n}{d t}&=\frac{D}{a^2}\left(L_{n+1}-2L_n +L_{n-1}\right)+\frac{k_{r}}{2}\left(S_n+S_{n+1}\right)
-kL_n+\frac{1}{2}\left(\nu_{n}+\nu_{n+1}\right)\label{teq:l2}\\
\frac{d S_n}{d t}&=\frac{k}{2}\left(L_{n-1}+L_{n}\right)-k_rS_n-k_oS_n\label{teq:se2}\,,\\
\frac{d S^{(i)}_n}{d t}&=k_{i}S_{n}-k_{2}S_{n}^{(i)}\label{teq:si}\,
\end{align}

In \cref{teq:si}, $k_{2}$ denotes the degradation rate in the immobile pool.
This immobile pool characterizes a third, non-diffusive, relaxation mode with dispersion relation 
\begin{equation}
s_3=k_2\,,
\end{equation}
with effective diffusion coefficient $D_3=0$ and effective degradation rate $K_3=k_2$.

\subsubsection{Application results and discussion}

\begin{figure}[]
\centering
\includegraphics[width=0.5\linewidth]{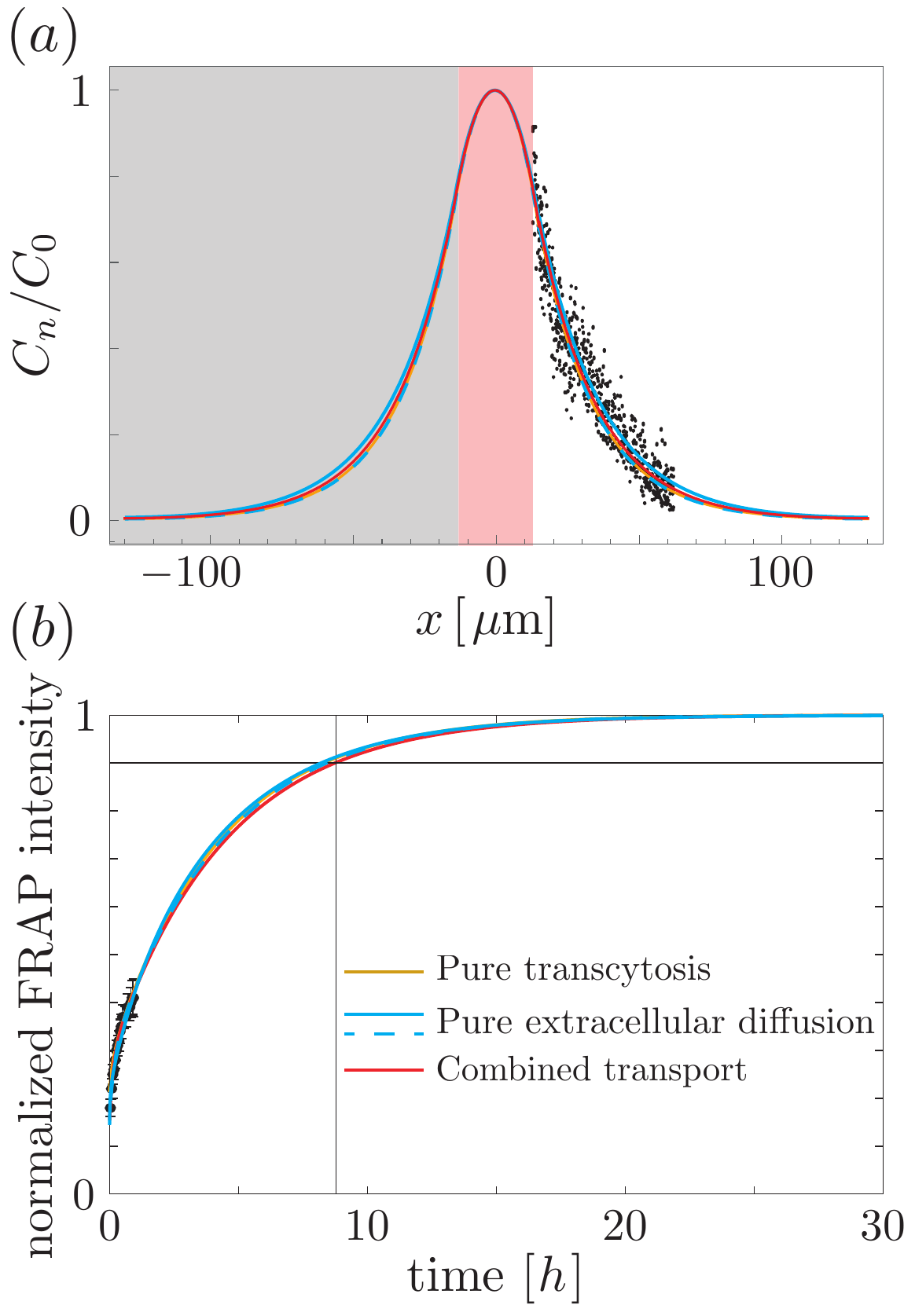}
\caption{\textbf{Comparison of steady state profiles and FRAP recovery curves between
experiment and transport model.}
(a)  Morphogen profile quantified in \cite{Kicheva:2007bha} (dots) shown together with 
calculated steady state profiles (solid and dashed lines) for the three transport regimes 
(color codes for the calculated profiles as in (b)). 
Red region indicates the source (width $(2w+1)a\approx10a$). The grey region corresponds to the
anterior part of the tissue.
(b)  Experimentally observed FRAP recovery \cite{Kicheva:2007bha}
(dots) shown together with calculated recovery curves for the three transport regimes. 
Vertical black line indicates $8h$ and horizontal black line, $90\%$ recovery. 
Experiments showed a $90\%$ recovery after about $8h$ in a pulse chase assay \cite{Entchev:2000tg}. 
Parameter values given in \cref{tab:params}.}
\label{fig:expss}
\end{figure}

We can now discuss experimental data on gradients of the morphogen Dpp in the developing wing imaginal disc of the fly.
Using a GFP-Dpp construct, the shape of the concentration profiles could be quantified for different stages of development \cite{Kicheva:2007bha}.
Quantification of the Dpp profile as a function of the distance of the anterior-posterior compartment boundary is shown in \cref{fig:expss}a (black dots), together with the profiles calculated for the four different transport scenarios 
(solid and dashed lines). They correspond to pure extracellular diffusion with small ($D=0.1\micron^2/s$, solid blue line)
and large ($D=20\micron^2/s$, dashed blue line) diffusion coefficient, pure transcytosis (solid yellow line) and
a combination of both (solid red line). Here we consider two different diffusion coefficients in the pure
extracellular diffusion scenario to be able to discuss different values suggested in the literature \cite{Zhou:2012ia}, 
see discussion below.
In all cases the 
decay length is about $\lambda\approx20\micron$. 

\begin{figure*}[]
\centering
\includegraphics[width=1\textwidth]{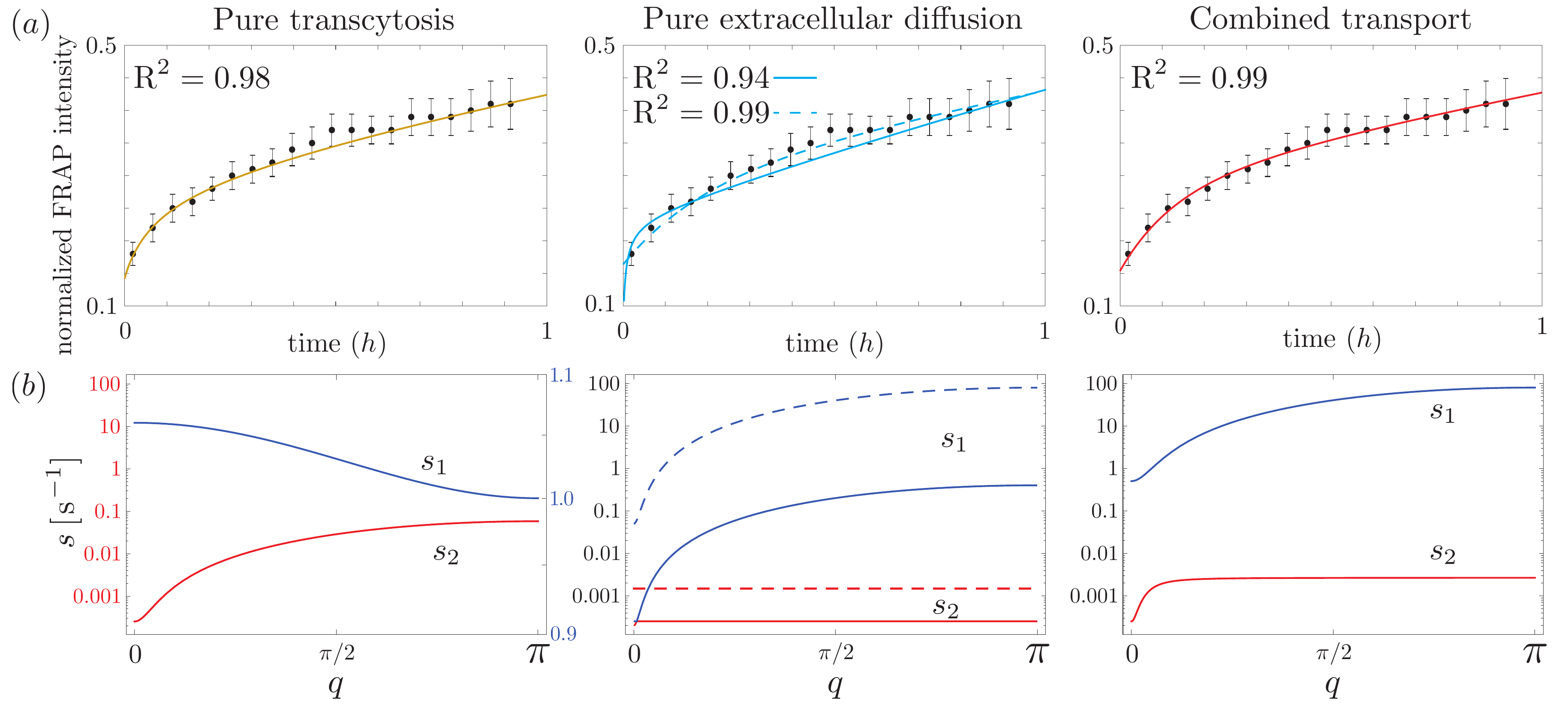}
\caption{\textbf{FRAP recovery curves and corresponding relaxation modes in three transport
scenarios.} 
(a) Calculated FRAP recovery curves as a function of time 
(solid and dashed lines) shown
together with experimental data \cite{Kicheva:2007bha}  
(dots, bars show standard error, 8 samples) for 
pure transcytosis scenario, pure extracellular diffusion scenario and a combined
transport scenario. For the pure extracellular diffusion scenario we 
show two simulations with slow (solid) and fast (dashed) diffusion coefficient. Parameters indicated in the figure correspond to the scenario with slow diffusion coefficient. The scenario with fast diffusion coefficient corresponds to $D=20\micron^2/s$ and $k=5.0\,10^{-2}\,\text{s}^{-1}$ \cite{Zhou:2012ia}. See full parameter sets in \cref{tab:params}.
(b) Dispersion relations of the relaxation rates as a function of wave-number $q$ corresponding to the
calculations shown in (a).
Parameter values are given in \cref{tab:params}. Parameters are determined to fit the average data with errors.
}
\label{fig:3regimes}
\end{figure*}

In order to determine kinetic parameters, fluorescence recovery after photobleaching (FRAP) was performed to quantify the recovery of the bleached fluorescence GFP-Dpp as a function of time.
The experimental data are shown in \cref{fig:expss}b  (black dots with error bars) together with calculated 
FRAP recovery curves for the four transport scenarios discussed above (solid and dashed lines).
All four scenarios are consistent with the experimental data shown in  \cref{fig:expss}a and b.
\Cref{fig:3regimes} shows the FRAP recovery within the first hour. The calculated recovery curves for the pure transcytosis, the pure extracellular diffusion and the combined transport scenarios are shown as fits to the 
experimental data, as solid and dashed lines, respectively. The values of the fit parameters as well as the values of the effective diffusion coefficient and effective degradation rate are shown for the four transport scenarios in \cref{tab:params}. 
These parameters are provided as single set estimates per case of study. Uncertainties are large as is reflected in the fact that for the experimental data available, we cannot discriminate between different transport scenarios.
The corresponding 
dispersion relations of transport modes as a function of the wave number $q$ are shown in \cref{fig:3regimes}b.
They are remarkably different, reflecting the properties of different transport mechanisms and yet they can
capture the same dynamics of FRAP recovery and the same steady state profile.

We find that the three transport regimes studied here agree with the full set of experimental data available, namely time-scale for gradient formation close to steady-state \cite{Entchev:2000tg}, the decay-length $\lambda$ and FRAP recovery curves \cite{Kicheva:2007bha}.
The effective diffusion coefficient $D_2$ and the effective degradation rate $K_2$ of the slower transport mode obtained for pure transcytosis and combined transport agree with the values estimated in \cite{Kicheva:2007bha}.
Future work adding further independent experimental assays will help determine full fits and confidence to the parameters presented here, and will help distinguishing between transport regimes.

%
%
%
%
\begin{table}
\caption{\label{tab:params}
List of parameter values for different transport scenarios considered: Pure transcytosis (yellow), Pure extracellular diffusion with slow (dashed blue) and fast (solid blue) diffusion coefficient, and Combined transport (red). Color references in parenthesis correspond to colors used in \cref{fig:expss} and \cref{fig:3regimes}a.
Parameters are: Extracellular diffusion coefficient $D$, internalization rate $k$, 
recycling rate $k_{r}$, output rate $k_{o}$ of molecules from the intracellular mobile pool, 
immobilization rate $k_{i}$ of molecules in the mobile pool, degradation rate  $k_{2}$ of molecules in the immobile 
pool, cell size $a$ and the bleaching depth $b$ as the fluorescence intensity at time $t=0$. 
Effective diffusion coefficients and effective degradation rates $D_{2},\,K_{2},\,D_{1},\,K_{1}$ calculated from \cref{teq:Ks,teq:Ds}, decay length $\lambda$ calculated from \cref{teq:lambdaapprox}. And R$^2$ of the fit to the FRAP data
\footnote{
\footnotesize{
Model parameter values are determine as follows: $D$ is taken from \cite{Kicheva:2007bha} and \cite{Zhou:2012ia}, $a$ is taken from \cite{Kicheva:2007bha}, $k_2$ is estimated from \cite{Entchev:2000tg} in Pure transcytosis and Combined transport cases. And $k$ is taken from \cite{Zhou:2012ia} for the case of Pure extracellular diffusion in the scenario of fast diffusion, and $b$ is chosen lower than the first recovery point. Two out of four remaining parameters are estimated from the values of $D_2$ and $K_2$ from reference \cite{Kicheva:2007bha}, and \cref{teq:lambdaapprox,teq:Ks,teq:Ds}. Note that we have given flexibility to these values within their standard deviation interval to heuristically find agreement to the experimental data. This process left two free parameter per case of study.
}
}
}
\begin{ruledtabular}
\begin{tabular}{ccccc}
&\mbox{Pure}&\multicolumn{2}{c}{\mbox{Pure extracellular}}&\mbox{Combined }\\
\mbox{Parameters}&\mbox{transcytosis}&\multicolumn{2}{c}{\mbox{diffusion}}&\mbox{transport}\\
\hline
$D\,[\micron^{2}/\second]$&0&0.10&20&20\\
$k\,[1/\second]$&0.5& $2.0\,10^{-4}$ &$5.0\,10^{-2}$& 0.5 \\
$k_{r}\,[1/\second]$  &$6.0\,10^{-2}$  & 0&0& $2.4\,10^{-3}$   \\
$k_{o}\,[1/\second]$  &$2.5\,10^{-4}$&$2.5\,10^{-4}$&$1.5\,10^{-3}$&$2.5\,10^{-4}$\\
$k_{i}\,[1/\second]$  &$2.5\,10^{-4}$&$1.25\,10^{-4}$&$3.0\,10^{-4}$&$2.5\,10^{-4}$\\
$k_{2}\,[1/\second]$  &$6.5\,10^{-5}$&$6.5\,10^{-5}$&$6.5\,10^{-5}$&$6.5\,10^{-5}$\\
$a\,[\micron]$ &$2.6$ &$2.6$ &$2.6$ &$2.6$\\
$b$&0.14&0.12&0.16&0.16\\
\hline
$D_{2}\,[\micron^{2}/\second]$&$0.09$&0.10&0&0.10\\
$K_{2}\,[1/\second]$&$2.2\,10^{-4}$& $2.0\,10^{-4}$ &$9.0\,10^{-4}$& $2.5\,10^{-4}$ \\
\hline
$D_{1}\,[\micron^{2}/\second]$&$-0.09$&0&20&20\\
$K_{1}\,[1/\second]$&$0.56$& $2.5\,10^{-4}$ &$5.0\,10^{-2}$& 0.5 \\
\hline
$D_{3}\,[\micron^{2}/\second]$&$0$&0&0&0\\
$K_{3}\,[1/\second]$&$6.5\,10^{-5}$&$1.25\,10^{-4}$&$1.8\,10^{-5}$&$6.5\,10^{-5}$\\
\hline
$\lambda\,[\micron]$&20.2& 22.4 &20.0& 21.0 \\
\hline
R$^2$&0.98&0.94&0.99&0.99\\
\end{tabular}
\end{ruledtabular}
\end{table}

\section{Conclusions}

We have presented a general cell-based framework for morphogen transport 
building on an earlier discrete model \cite{Bollenbach:2007vf}, where we discussed the implication of directional bias in morphogen transport and its effect on explicit ligand-receptor dynamics. Here we bring the concept of eigen-modes to the problem of gradient formation, and studied the mode
structure of such a model revealing emergent long wavelength behaviors that cannot be captured
by continuum models. Within a common framework, this allows us to study 
  extreme models of morphogen transport 
  that have been debated in the literature.
The main controversial point of discussion was whether the shape of the Dpp gradient is solely
set by a combination of rapid extracellular diffusion and terminal uptake by cells (pure extracellular
diffusion, \cite{Lander:2002cq}) or whether 
uptaken molecules can return to the extracellular space and contribute to the formation of the
gradient profile (combined transport \cite{Entchev:2000tg}). 
In an extreme limit (pure transcytosis scenario) the molecules do not diffuse
extracellularly and are transferred directly from cell to cell.

Our transport model exhibits two relaxation modes, one fast and one slow,
characterized by wave-length dependent dispersion relations.
For the slow transport mode we can define an effective diffusion coefficient and an effective degradation rate
which govern the large scale dynamics of the concentration profile. These effective transport parameters
set the decay length of gradients in steady state and they capture the long time-scale dynamics in experiments such
as FRAP in the tissue. In contrast, measurements of extracellular diffusion by FCS provide information
about one of the parameters in the model, the extracellular diffusion coefficient $D$. These two parameters
(the effective diffusion coefficient $D_2$ and the extracellular diffusion coefficient $D$) are conceptually
different and can indeed differ significantly in value. For example, in the combined transport scenario
$D_2\simeq 0.1\mu m^2/s$ and $D=20\mu m^2/s$, see Fig. 5 and table I. This could account for
apparent discrepancies between different types of experiments such as FRAP \cite{Kicheva:2007bha} and 
FCS  \cite{Zhou:2012ia} that has led to controversies in the field.

This work provides a new framework based on hydrodynamic modes of transport within which to study the dynamics of morphogen gradients and can be used, together with experimental assays, to bridge gaps and inconsistencies in the field. We discussed how to use different experimental assays (spatial concentration profiles, FRAP, FCS and long time relaxation of morphogen gradient) to estimate values for trafficking parameters. Further transport details can be discussed within the same framework by extending the model, see Apendix \ref{Ap:comp5}, which will require additional independent experimental assays.

\begin{acknowledgments}
D.A.H. thanks Marko Popovic, S\'andalo Rold\'an-Vargas and Johanna Dickmann for fruitful discussions. D.A.H., F.J. and M.G.G. acknowledge support from the DIP of the Canton of Geneva, SNSF, the SystemsX epiPhysX grant, the ERC (Sara and Morphogen), the NCCR Chemical Biology program and the Polish-Swiss research program. Z.H. was supported by an HFSP Long Term Fellowship. This research was supported in part by the National Science Foundation under Grant No. NSF PHY-1748958, NIH Grant No. R25GM067110, and the Gordon and Betty Moore Foundation Grant No. 2919.01.
\end{acknowledgments}

\appendix

\section{Steady state solutions to the transport equations}\label{Ap:ss}

The steady state solution of \cref{teq:l,teq:se} for a source with constant production rate $\nu_{n}=\nu$ for
$-w\leq n \leq w$ and $\nu_{n}=0$ outside the
source region, 
has the form
\begin{equation}\label{eq:ssww}
\begin{pmatrix}
L_{n}^{ss}\\
S_{n}^{ss}
\end{pmatrix}
=
\begin{pmatrix}
L_{}^{0}\\
S_{}^{0}
\end{pmatrix}
+
\begin{pmatrix}
L_{}^{-}\\
S_{}^{-}
\end{pmatrix}
e^{-n\sigma}
+
\begin{pmatrix}
L_{}^{+}\\
S_{}^{+}
\end{pmatrix}
e^{n\sigma}\,,
\end{equation}
with nonvanishing $(L_{}^{0},S_{}^{0})$ in the source region. Here  $(L_{}^{\pm},S_{}^{\pm})$ are the amplitude of the positive and negative exponential contributions to the spatial concentration profile, respectively. 
We distinguish 
three regions (i) from $n=-N$ to $n=-w$ with amplitudes $(L_1,S_1)$ (ii) from $n=-w$ to $n=w$ with
amplitudes $(L_2,S_2)$ and (iii) from $n=w$ to $n=N$ with
amplitudes $(L_3,S_3)$. The amplitudes $L^0_i,\,L^-_i,L^+_i$  for $i=1,2,3$ are given by:

\begin{align}
L^0_1=&0\label{eq:L10}\\
L^-_1=&L_2^0\gamma^{-1}   \left(e^{ (2   w+1)\sigma}-1\right) e^{ -(w+1)\sigma}\left(\chi e^{\sigma }+\tilde{\chi}\right)\label{eq:L11}\\
L^+_1=&L_2^0\gamma^{-1}  \left(e^{(2 w+1)\sigma }-1\right) e^{(2 N -  w)\sigma} \left(-\chi-\tilde{\chi} e^{\sigma }\right)\label{eq:L12}\\
L^0_2=&L_2^0\gamma^{-1}   \frac{\nu+2(L_2^-+L_2^+) (-\xi+\tilde{\chi}+\chi \cosh (\sigma ))}{-\xi+\chi+\tilde{\chi}}\label{eq:L20}\\
L^-_2=&L_2^0\gamma^{-1} \left( \chi \cosh ( (N-w-1)\sigma )+ \tilde{\chi}\cosh ( (N-w)\sigma )\right)  2e^{N \sigma } \label{eq:L21} \\
L^+_2=&L_2^0\gamma^{-1}  ( \chi \cosh (  (N-w-1)\sigma)+\tilde{\chi} \cosh ( (N-w)\sigma ))  2e^{(N+1) \sigma } \label{eq:L22} \\
L^0_3=&0 \label{eq:L30}\\
L^-_3=&L_2^0\gamma^{-1}  \left(e^{ (2   w+1)\sigma}-1\right) e^{ (2 N-w-1)\sigma } \left(-\chi-\tilde{\chi} e^{\sigma }\right) \label{eq:L31}\\
L^+_3=&L_2^0\gamma^{-1}   \left(e^{(2  w+1)\sigma }-1\right) e^{-w \sigma}   \left(\chi e^{\sigma }+\tilde{\chi}\right) \label{eq:L32}
\end{align}
with $\gamma=-2 \left( \chi \left( \text{exp}(2 N \sigma )+ \text{exp}(\sigma )\right)+\tilde{\chi} \left( 1+\text{exp}((2 N +1)\sigma )\right)\right)$, $\chi=D/a^2+k k_r/(4(k_r+k_o))$, $\tilde{\chi}=-D/a^2+k k_r/(4(k_r+k_o))-k/2$, $\xi=k/2$ and decay rate $\sigma$ as in \cref{teq:coshsigma}.

The amplitudes of the steady state solution for the intracellular pool are proportional to 
the amplitudes of the extracellular pool. From \cref{teq:se}, we have
\begin{equation}
S_n=\frac{1}{2}\frac{k}{k_r+k_o}\left(L_{n-1}+L_n \right).
\end{equation}
The amplitudes $S_i^0,S_i^-,S_i^+$ for $i=1,2,3$ read:
\begin{align}
S^0_i&=\Omega L^0_i\\
S^-_i&=\Omega L^-_i\\
S^+_i&=\Omega L^+_i.
\end{align}
with $\Omega=k\left(e^{-\sigma}+1\right)/(2k_r+2k_o)$.

\section{Five-compartment model of morphogen transport}\label{Ap:comp5}

In our approach, it is straightforward to consider additional phenomena in the transport process. 
Here, we define five pools of molecules, see \cref{fig:scheme,eq:l,eq:sr,eq:sl,eq:se,eq:si}. 
We denote by $\bar{L}_n$ the number of  
extracellular ligand molecules located between cell $n$ and $n+1$ and not bound to receptors.
The number of ligand molecules bound to receptors at the 
plasma membrane at the right and left sides of cell $n$ are denoted
$S_n^{(r)}$ and $S_n^{(\ell)}$, respectively. The number of ligand molecules internalized upon receptor binding
is denoted $S_n^{(e)}$. Molecules in this pool could be degraded, recycled back to the plasma
membrane or transferred to a pool of intracellular molecules whose number is
denoted  $S_n^{(i)}$. The $S_n^{(i)}$ pool can be degraded but does not return
to the $S_n^{(e)}$ pool.

\subsection{Dynamic equations}

The dynamic equations of the five compartment model read:
 \begin{figure}[b]
\centering
\includegraphics[width=1\linewidth]{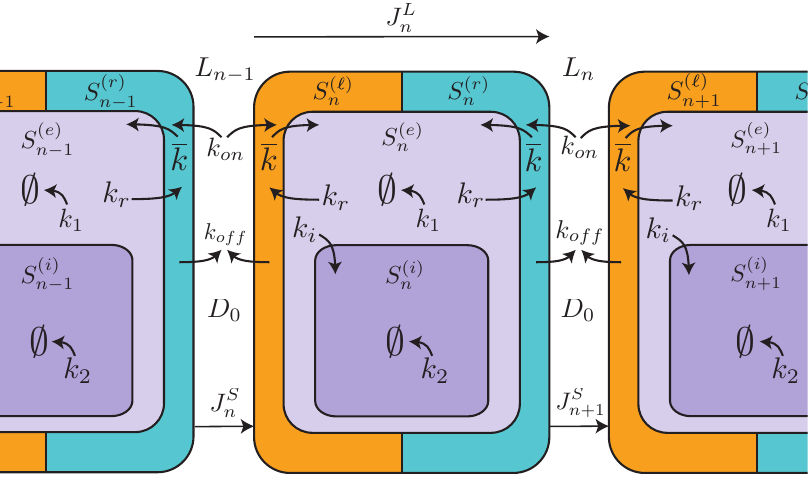}
\caption{\textbf{Scheme of the morphogen transport model with five cellular compartments} 
We introduce pools of receptor bound ligands at the left and right surface of cell $n$. The molecule
numbers in these pools are denoted $S^{(\ell)}$ and $S^{(r)}$, respectively. The number of
free extracellular ligand molecules which diffuse with diffusion coefficient $D_0$
are denoted $\bar L_n$. The molecule numbers in the intracellular
mobile and immobile pools are $S^{(e)}$ and $S^{(i)}$, respectively.
The binding rate of free ligand to cell surfaces is denoted $\bar k$, $k_r$
is the recycling rate of receptor bound ligands in the mobile pool to the cell surface,
molecules of this mobile pool are
 degraded with degradation rate $k_1$, or transfer to the immobile pool with rate $k_i$. 
 Degradation rate of molecules in the immobile pool is denoted 
 $k_2$. The transport fluxes $J_n^L$ and $J_n^S$ are defined in \cref{eq:fluxnL,eq:fluxnS}.}
\label{fig:scheme}
\end{figure}
\begin{align}
\frac{d \bar{L}_n}{d t}&=\frac{D_0}{a^2}\left(\bar{L}_{n-1}-2\bar{L}_n +\bar{L}_{n+1}\right)+k_{off}\left(S_n^{(r)}+S_{n+1}^{(\ell)} \right)
-k_{on}\bar{L}_n+\frac{1}{2}(\nu_{n}+\nu_{n+1})\label{eq:l}\\
\frac{d S_n^{(r)}}{d t}&=-(k_{off}+\bar{k})S_n^{(r)}+\frac{k_{on}}{2}\bar{L}_n+\frac{k_r}{2}S_n^{(e)}\label{eq:sr}\\
\frac{d S_n^{(\ell)}}{d t}&=-(k_{off}+\bar{k})S_n^{(\ell)}+\frac{k_{on}}{2}\bar{L}_{n-1}+\frac{k_r}{2}S_n^{(e)}\label{eq:sl}\\
\frac{d S_n^{(e)}}{d t}&=\bar{k}\left(S_n^{(\ell)}+S_n^{(r)} \right)-k_rS_n^{(e)}-(k_1+k_i)S_n^{(e)}\label{eq:se}\\
\frac{d S_n^{(i)}}{d t}&=k_iS_n^{(e)}-k_2S_n^{(i)}\label{eq:si}\, .
\end{align}
These apply for $\bar L_n$ if $-N-1\leq  n \leq N$ and for $S^{(r,\ell,e,i)}_n$ if $-N\leq  n \leq N$.
Morphogen molecules are produced and secreted to contribute to the $\bar{L}$ pool in the extracellular space
with rate $\nu_n$.
Molecules of the $\bar L_n$ pool diffuse with diffusion coefficient $D_0$. 
They can bind to receptors on the plasma membrane 
at the left and right side of cells with binding rate $k_{on}$. 
Bound ligand can unbind with rate $k_{off}$ and can be internalized with 
rate $\bar{k}$ into the pool $S^{(e)}_n$. Molecules of the pool $S^{(e)}_n$ can recycle back with rate $k_r$ to the 
plasma membrane, they can be degraded with degradation rate $k_1$ or they can be
transferred with rate $k_i$ to the immobile pool $S_n^{(i)}$. Finally, the immobile pool is degraded at rate $k_2$. 
For simplicity we choose the
boundary conditions as $S_{-N-1}^{(r,e,i)}=S^{(\ell,e,i)}_{-N}$, $S_{N}^{(r,e,i)}=S^{(\ell,e,i)}_{N+1}$
and $\bar L_{-N-2}=\bar L_{-N}$ and $\bar L_{N+1}=\bar L_{N-1}$.

We define the total number of morphogen molecules per cell 
\begin{equation}\label{eq:totcon}
C_{n}=\frac{1}{2}\left(\bar{L}_{n-1}+\bar{L}_{n} \right)+S^{(r)}_{n}+S^{(l)}_{n}+S^{(e)}_{n}+S^{(i)}_{n}\,.
\end{equation}
and the currents $J_n^{L},\,J_n^{S}$. 
The current of ligands $J_n^{L}$ describes the transport of ligand across cells 
via extracellular diffusion with coefficient $D$.
 The current  $J_n^{S}$ describes transport between cells via binding and 
 unbinding of molecules from receptors on the plasma membrane. These currents read
\begin{align}\label{eq:fluxnL}
J_n^{L}=&-\frac{D_0}{a^{2}}\left(\bar{L}_{n}-\bar{L}_{n-1}\right)\\
J_n^{S}=&-k_{off}\left(S_{n}^{(\ell)}-S_{n-1}^{(r)}\right)\,.\label{eq:fluxnS}
\end{align}
The balance equation for total molecule number then reads
\begin{align}\label{eq:balance}
\frac{dC_n}{dt}=&\frac{1}{2}(J_{n-1}^{L}-J_{n+1}^{L})+\frac{1}{2}(J_{n}^{S}-J_{n+1}^{S})
-k_{1}S_{n}^{(e)}-k_{2}S_{n}^{(i)}+\frac{1}{4}\left(\nu_{n-1}+2\nu_{n}+\nu_{n+1}\right)\,.
\end{align}

\subsection{Dynamic modes of transport}

The general solution to the dynamic \cref{eq:l,eq:sr,eq:sl,eq:se,eq:si} can 
be written as
\begin{equation}\label{eq:sspluspertD}
\mathbf{c}_n(t)
=
\mathbf{c}_n^{ss}
+
\sum_{\alpha=1}^5
\sum_{m=-N}^N
a_{m}^{\alpha}
\mathbf{c}_{m}^{\alpha}
e^{iq_{m}n}
e^{-s_{\alpha}(q_{m})t}\,.
\end{equation}
with concentration vector  $\mathbf{c}_n=(\bar{L}_n, S^{(r)}_n, S^{(l)}_n, S^{(e)}_n, S^{(i)})$.
Here,
the time-independent term corresponds to the steady state profile $\mathbf{c}_n^{ss}$. 
The time-dependent
terms are relaxation modes of wave number $q_{m}$
and relaxation rate $s_{\alpha}$, where $\alpha=1,\dots, 5$ 
is a mode index. The mode amplitudes are denoted
$a_{m}^{\alpha}$. The boundary conditions are consistent with wave numbers
\begin{equation}\label{eq:Fq}
q_{m}=\frac{2\pi m}{2N+1}\, .
\end{equation}

\begin{figure}[h]
\centering
\includegraphics[width=1\linewidth]{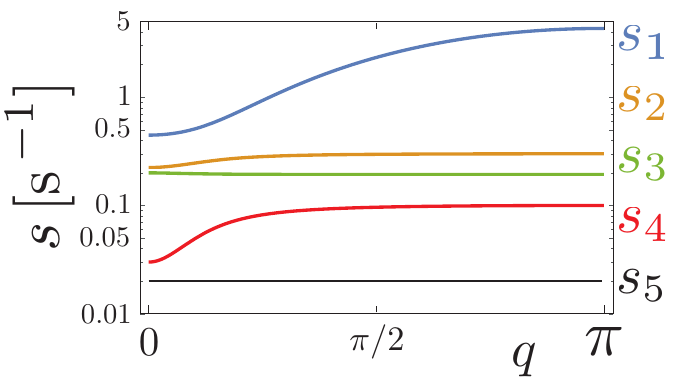}
\caption{\textbf{Example of the dispersion relations of the five compartment model.} 
Shown are the relaxation rates $s_\alpha$ as a function of wave number $q$ for
the five relaxation modes $\alpha=1,\dots ,5$ of the full transport model.
Parameter values: $D_{0}=10\mu\text{m}^{2}/\text{s},\,k_{on}=0.3/\text{s},\,k_{off}=0.1/\text{s},\,\bar{k}=0.1/\text{s},\,k_{r}=0.1/\text{s},\, k_{o}=0.1/\text{s},\,k_{2}=0.02/\text{s},\,a=1\mu\text{m}$.}
\label{fig:disp_rel}
\end{figure}

The relaxation rates and the
mode eigenvectors $\mathbf{c}_m^{\alpha}$
follow from an eigenvalue problem:
\begin{equation}\label{eq:eigenprob}
M(iq_{m})\,
\mathbf{c}_{m}^{\alpha}
=
-s_{\alpha}
\mathbf{c}_{m}^{\alpha}\,.
\end{equation}
Here $M(z)$ with $z=iq_{m}$ is the matrix 
\begin{widetext}
\begin{equation}\label{eq:M}
M=\begin{pmatrix}
-k_{on} + D_0/a^{2}\left(e^{-z}-2+e^{z} \right)&k_{off}&k_{off}e^{z}&0&0\\
k_{on}/2&-k_{off}-\bar{k}&0&k_r/2&0\\
k_{on}e^{-z}/2&0&-k_{off}-\bar k&k_r/2&0\\
0&\bar{k}&\bar{k}&-k_r-k_o&0\\
0&0&0&k_i&-k_2
\end{pmatrix}\,,
\end{equation}
\end{widetext}
The eigenvalue problem \cref{eq:eigenprob} defines a characteristic polynomial, 
$\det[M+\mathbb{I}s]=0$, where $\mathbb{I}$ is the identity matrix, which reads
\begin{align}\label{eq:cpoly5eqs}
0&=(k_2-s)\Big[(k_{off}+\bar k- s)\Big(\bar k(k_o- s)(k_{on}- s)
- s(k_{r}+k_o- s)(k_{off}+k_{on}- s) \Big)   \nonumber \\ 
&+  (\cosh{z}-1) \Big(  \frac{1}{2}  \bar{k} k_{off} k_{on} k_r +      \frac{2 D_0}{a^2} (k_{off}+\bar{k}  - s) (\bar{k} (k_o - s) + (k_{off} - s) (k_o + k_r - s))\Big) \Big]\,. 
\end{align}
This equation defines a fifth order polynomial equation in $s$ which has five zeros that 
of wave-length of the eigenmodes of the system. The corresponding eigenvectors 
$\mathbf{c}_m^{\alpha}$ follow from \cref{eq:eigenprob}.
\cref{fig:disp_rel} shows an example of the dispersion relations of the five relaxation modes.
\subsection{Steady-state concentration profiles}

\Cref{eq:l,eq:sr,eq:sl,eq:se,eq:si} can be solved in a piecewise manner for regions of constant 
production rate. The solution reads
\begin{equation}
\mathbf{c}_n^{ss}= \begin{cases} 
      \mathbf{c}_{1}^-e^{-\sigma n}+\mathbf{c}_{1}^+e^{\sigma n} & -N\leq n\leq -w \\
     \mathbf{c}_{2}^0+\mathbf{c}_{2}^-e^{-\sigma n}+\mathbf{c}_{2}^+e^{\sigma n} & -w\leq n\leq w \\
      \mathbf{c}_{3}^-e^{-\sigma n}+\mathbf{c}_{3}^+e^{\sigma n} & w\leq n\leq N 
   \end{cases}
\end{equation}
 with concentration vector in steady state $\mathbf{c}_n^{ss}=(L_n\,S^{(r)}_n\,S^{(\ell)}_n\,S^{(e)}_n\,S^{(i)})$.
Amplitudes $\mathbf{c}^{(0,-,+)}_{i}$ for $i=1,2,3$ are 
obtained applying boundary conditions as defined above 
and matching conditions at the boundaries between the three regions. 
The amplitudes can be written in the form 
\begin{equation}
\mathbf{c}^{(0,-,+)}_{i}=A_i^{(0,-,+)}\mathbf{V}
\end{equation}
for $i=1,2,3$, where $\mathbf{V}=[v_{1},\,v_{2},v_{3},\,v_{4},\,v_{5}]$ is a vector with
components
\begin{align}
v_1^0=&1\,,\\
v_2^0=&\frac{1}{2}\frac{k_r}{\bar{k}+k_{off}}+  \frac{1}{1+e^{-a\sigma}}\left(\frac{k_o}{\bar{k}}+\frac{k_r}{\bar{k}}\frac{1}{1+\bar{k}/k_{off}}\right)v_4^0\,,\\
v_3^0=&\frac{1}{2}\frac{k_r}{\bar{k}+k_{off}}+  \frac{e^{-a\sigma}}{1+e^{-a\sigma}}\left(\frac{k_o}{\bar{k}}+\frac{k_r}{\bar{k}}\frac{1}{1+\bar{k}/k_{off}}\right)v_4^0\,,\\
v_4^0=&\frac{1}{2}\frac{k_{on}(1 + e^{-a\sigma}) }{k_o+ k_{off}/\bar{k}(k_r+k_o)}\,,\\
v_5^0=&\frac{k_i}{k_2}v_4^0\,.
\end{align}
The coefficients $A_i^{(0,-,+)}$ are given by
\begin{align}
A^0_1=&0\label{eq:A10}\\
A^-_1=&A_2^0\gamma^{-1}   \left(e^{ (2   w+1)\sigma}-1\right) e^{ -(w+1)\sigma}\left(\chi e^{\sigma }+\tilde{\chi}\right)\label{eq:A11}\\
A^+_1=&A_2^0\gamma^{-1}  \left(e^{(2 w+1)\sigma }-1\right) e^{(2 N -  w)\sigma} \left(-\chi-\tilde{\chi} e^{\sigma }\right)\label{eq:A12}\\
A^0_2=&A_2^0\gamma^{-1}   \frac{\nu+2(A_2^-+A_2^+) (-\xi+\tilde{\chi}+\chi \cosh (\sigma ))}{-\xi+\chi+\tilde{\chi}}\label{eq:A20}\\
A^-_2=&A_2^0\gamma^{-1} \left( \chi \cosh ( (N-w-1)\sigma )+ \tilde{\chi}\cosh ( (N-w)\sigma )\right)  2e^{N \sigma } \label{eq:A21} \\
A^+_2=&A_2^0\gamma^{-1}  ( \chi \cosh (  (N-w-1)\sigma)+\tilde{\chi} \cosh ( (N-w)\sigma ))  2e^{(N+1) \sigma } \label{eq:A22} \\
A^0_3=&0 \label{eq:A30}\\
A^-_3=&A_2^0\gamma^{-1}  \left(e^{ (2   w+1)\sigma}-1\right) e^{ (2 N-w-1)\sigma } \left(-\chi-\tilde{\chi} e^{\sigma }\right) \label{eq:A31}\\
A^+_3=&A_2^0\gamma^{-1}   \left(e^{(2  w+1)\sigma }-1\right) e^{-w \sigma}   \left(\chi e^{\sigma }+\tilde{\chi}\right) \label{eq:A32}
\end{align}
with
$\gamma=-2 ( \chi ( \text{exp}(2 N \sigma )+ 
\text{exp}(\sigma ))+\tilde{\chi} \left( 1+\text{exp}((2 N +1)\sigma )\right))$, 
$\chi=D_0/a^2+k_r k_{off}/(2(k_{off}+\bar{k}))\Gamma$, 
$\tilde{\chi}=-D_0/a^2+ k_{off}/(2(k_{off}+\bar{k}))(k_{on}+k_r \Gamma)-k_{on}/2$, $\xi=k_{on}/2$, 
$\Gamma=k_{on}\bar{k}/(2((k_r+k_i+k_1)(k_{off}+\bar{k})-k_r\bar{k}))$. The decay rate $\sigma$ is determined from 
the condition $\det(M(z=\sigma))=0$, which holds for steady states. 
We then find
\begin{equation}\label{eq:coshsigma}
\cosh{\sigma}=\frac{1}{2}\left[\frac{k_r}{4k_o}\frac{k_{off}}{(k_{off}+\bar{k})}+\frac{D_0}{a^2k_{on}}\left(1+\frac{k_{off}}{\bar{k}}\left(1+\frac{k_r}{k_o}\right)\right) \right]^{-1}+1\,.
\end{equation}
with $k_o=k_i+k_1$. 

The decay length of the graded distribution of molecules outside of the source is given by
\begin{equation}\label{eq:lambdasigma}
\lambda=\frac{a}{\sigma}\,.
\end{equation}
In the limit of large $\lambda\gg a$, the decay length can be approximated as
\begin{equation}\label{eq:lambdaapprox}
\lambda\approx a \left[\frac{k_r}{4k_o}\frac{k_{off}}{(k_{off}+\bar{k})}+\frac{D_0}{a^2k_{on}}\left(1+\frac{k_{off}}{\bar{k}}\left(1+\frac{k_r}{k_o}\right)\right)\right]^{1/2}\,.
\end{equation}
The decay length $\lambda$ again has two contributions, one dominated by recycling of molecules at 
rate $k_{r}$ and one dominated by extracellular diffusion with diffusion coefficient $D_0$. 
Interestingly, the unbinding rate $k_{off}$ plays an important role for the decay length.

\subsection{Effective diffusion constant and effective degradation rate}
We determine effective degradation rates $K_\alpha$ and effective diffusion coefficients 
$D_\alpha$ by expanding  the dispersion 
relations to second order in wave-length
\begin{equation}
s_\alpha\simeq K_\alpha + D_\alpha q^2
\end{equation}
\subsubsection{Effective degradation rates}

The effective degradation rates are given by $K_\alpha=s_\alpha(q=0)$.  The characteristic polynomial \cref{eq:cpoly5eqs} reads for $q=0$:
\begin{equation}
0=(k_{2}- s)(k_{off}+\bar k- s)\Big(\bar k(k_o- s)(k_{on}- s)
- s(k_{r}+k_o- s)(k_{off}+k_{on}- s) \Big)
\end{equation}
which has five zeros. We can identify $ s_2=k_{off}+\bar k$ and $s_5=k_{2}$.
The remaining three zeros can be obtained from 
\begin{equation}\label{eq:cpol3}
\bar k(k_o- s)(k_{on}- s)- s(k_{r}+k_o- s)(k_{off}+k_{on}- s)=0\,.
\end{equation}
\Cref{eq:cpol3} is a cubic equation
\begin{equation}\label{eq:pol3}
 s^3+a_2 s^2+a_1 s+a_0=0\,,
\end{equation}
where 
\begin{align}
a_2=&-\left(k_{r}+k_o+\bar k+k_{off}+k_{on}\right)\\
a_1=&(k_{r}+k_o)(k_{off}+k_{on})+\bar k(k_o+k_{on})\\
a_0=&-\bar k k_o k_{on}\,.
\end{align}
We define the discriminant of the cubic function as:
\begin{equation}\label{eq:pol3}
\Delta=\frac{1}{27}\left(4\Delta_0^3-\Delta_1^2 \right)\,,
\end{equation}
\noindent with 
\begin{align}
\Delta_0=&a_2^2-3a_1\\
\Delta_1=&2a_2^3-9a_2a_1+27a_0\,.
\end{align}
The discriminant is always positive, $\Delta>0$, 
which indicates that the cubic function has three real roots. 
We finally express the effective degradation rates:
\begin{align}
 K_1=&-\frac{1}{3}\left(a_2-2\sqrt{\Delta_0}\cos{\left(\frac{1}{3}\arccos{-\frac{\Delta_1}{2\sqrt{\Delta_0}}}\right)} \right)\label{eq:aK1}\\
  K_2=&k_{off}+\bar k\label{eq:aK2}\\
 K_3=&-\frac{1}{3}\left(a_2-2\sqrt{\Delta_0}\cos{\left(\frac{1}{3}\arccos{-\frac{\Delta_1}{2\sqrt{\Delta_0}}}-\frac{2\pi}{3}\right)} \right)\label{eq:aK3}\\
 K_4=&-\frac{1}{3}\left(a_2-2\sqrt{\Delta_0}\cos{\left(\frac{1}{3}\arccos{-\frac{\Delta_1}{2\sqrt{\Delta_0}}}-\frac{4\pi}{3}\right)} \right)\label{eq:aK4}\\
 K_5=&k_2\label{eq:aK5}\,.
\end{align}
We have chosen the order or rates from fast to slow as in \cref{fig:disp_rel}.
\subsubsection{Effective diffusion coefficients}\label{ap:effidiff}
In order to calculate the effective diffusion coefficient for the different modes, we consider the 
characteristic polynomial for $z=iq$ which has the form
\begin{equation}\label{eq:pol4}
(k_{2}-s) (s^4+b_3 s^3+b_2 s^2+b_1 s+b_0)=0\,.
\end{equation}
Here
\begin{small}\begin{align}
b_3=&-\left(k_{r}+k_o+2(k_{off}+\bar k)+k_{on} \right)- D_0q^2 \label{eq:b3}\\
b_2=&(k_{off}+\bar k)^2+(k_{r}+k_o)(\bar k+2k_{off})
+k_{on}\left((k_{r}+k_o)+k_{off}+2\bar k \right)+\bar k k_o\nonumber \\&
+ D_0q^2\left(k_{r}+k_o+2(k_{off}+\bar k) \right)\label{eq:b2}\\
b_1=&-\Big(k_{on}\big((k_{off}+\bar k)(\bar k+k_{o}+k_{r})+\bar kk_{o}\big)
 +(k_{off}+\bar k)((k_{off}+\bar k)k_{o}+k_{off}k_{r})\Big)\nonumber \\&
- D_0q^2\Big((k_{off}+\bar k)^2+2(k_{off}+\bar k)(k_{o}+k_{r})-\bar kk_o \Big)\label{eq:b1}\\
b_0=&(k_{off}+\bar k)\bar kk_ok_{on}
+q^2\Big(\frac{a^{2}}{4}\bar kk_{r}k_{on}k_{off}
+ D_0(k_{off}+\bar k)\big((k_{off}+\bar k)k_{o}+k_{off}k_r \big)\Big).\label{eq:b0}
\end{align}\end{small}
A simple way to solve \cref{eq:pol4} this is to rewrite it as a function of its solutions
\begin{equation}\label{eq:pol4fac}
( s- s_1)( s- s_2)( s- s_3)( s- s_4)(s-k_{2})=0\,,
\end{equation}
 \noindent and expand it in a four degree polynomial to identify coefficients with \cref{eq:b3,eq:b2,eq:b1,eq:b0}
 \begin{align}
b_3=&- s_1- s_2- s_3- s_4\label{eq:b3w}\\
b_2=& s_1( s_2+ s_3)+ s_2( s_3+ s_4)+ s_4( s_1+ s_3)\label{eq:b2w}\\
b_1=&- s_1 s_2( s_3+ s_4)- s_3 s_4( s_1+ s_2)\label{eq:b1w}\\
b_0=& s_1 s_2 s_3 s_4\label{eq:b0w}\,.
\end{align}
 From \cref{eq:pol4}, we observe that the mode with effective degradation $K_5=k_2$ does not depend on $q$, thus this is a non-diffusive mode and 
\begin{equation}\label{eq:aD5}
D_5=0\,.
\end{equation}
We then expand \cref{eq:b3w,eq:b2w,eq:b1w,eq:b0w} in power series of $q$ up to second order and identify the effective diffusion coefficients as the coefficient of the $q^2$-term as a function of the degradation rates $K_1,\, K_2,\, K_3,\, K_4$,  and the coefficients $b_0,\, b_1,\, b_2,\, b_3$ defined in \cref{eq:b3,eq:b2,eq:b1,eq:b0}, with \cref{eq:seriesqsmallq}. This process leads to the diffusion coefficients:
\begin{align}\label{eq:aallDs}
 D_\alpha=&\frac{-\sigma_0+ D_0\left( K_\alpha^3-\sigma_1 K_\alpha^2+\sigma_2 K_\alpha-\sigma_3 \right)}{\displaystyle \prod\limits_{\substack{\beta=1\\\beta\neq \alpha}}^4( K_\alpha- K_\beta)},\,\\&\text{for }\alpha=1,\,2,\,3,\,4\nonumber
\end{align}
\noindent with
\begin{align}
\sigma_0=&\frac{a^{2}}{4}k_{r}\bar kk_{on}k_{off}\\
\sigma_1=&k_{r}+k_o+2(k_{off}+\bar k)\\
\sigma_2=&(k_{off}+\bar k)(\bar k+k_{off}+2k_{o})+(2k_{off}+\bar k)k_{r}\\
\sigma_3=&(k_{off}+\bar k)\left((k_{r}+k_o)k_{off}+\bar kk_o \right)
\end{align}
Note that $K_i^3-\sigma_1 K_i^2+\sigma_2 K_i-\sigma_3 =0$ for $i=2$, thus the effective diffusion coefficient of the second mode, does not depend of the free diffusion coefficient, $D_{2}\neq D_{2}(D_{0})$, but depends on trafficking parameters.
\subsection{Dispersion relations in the complex plane}
\begin{figure}[h!]
\centering
\includegraphics[width=1\linewidth]{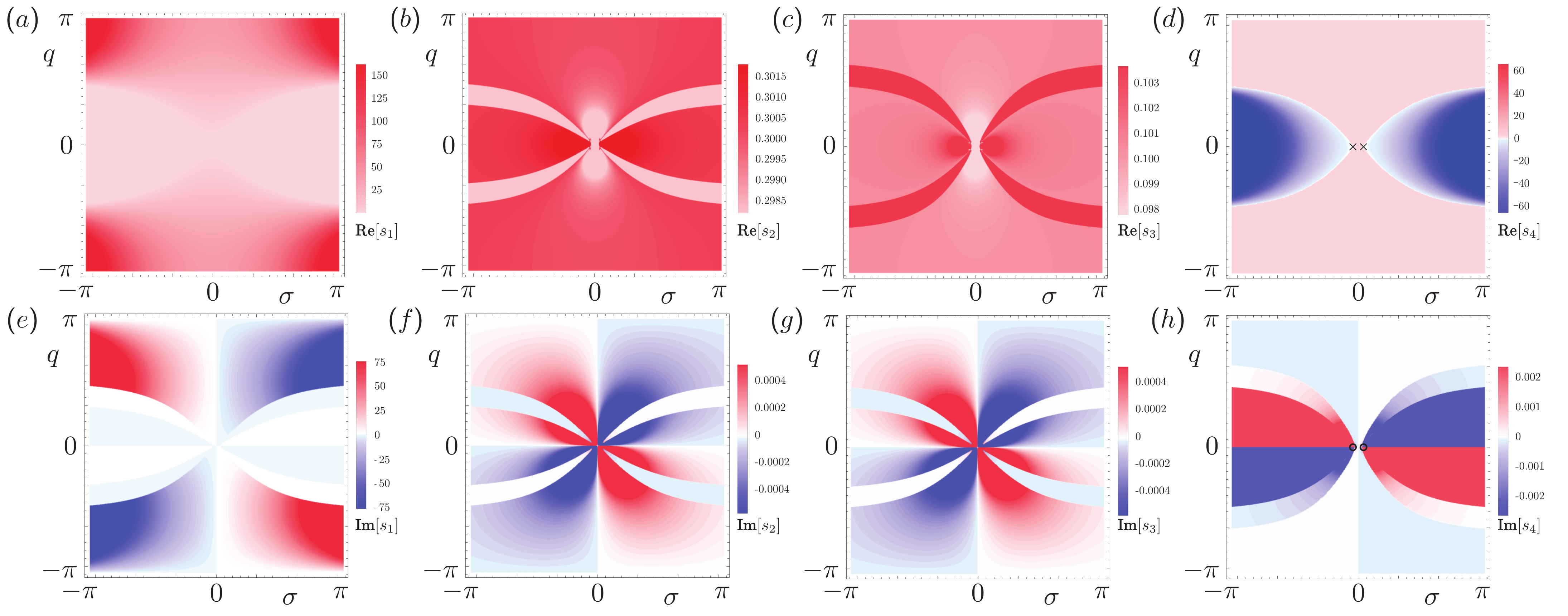}
\caption{\textbf{Dispersion relation of the five-compartment model in the complex plane.} 
(a)-(e) Real parts of the relaxation rates $s_\alpha$ with $\alpha=1,\dots, 4$
as a function of complex wave number $z=\sigma+iq$. (e)-(h)
Imaginary parts of the same relaxation rates as in (a)-(e).
Values of $z$ for which $s_4=0$ are indicated in (d) and (h) by black crosses. Parameter values 
as in \cref{fig:disp_rel} }
\label{fig:spectrum5}
\end{figure}

To discuss the timescale at which the shape of morphogen gradients is formed during its formation in a five-compartment transport model, we have analyzed the relaxation time spectrum in the complex plane from the zeros of \cref{eq:cpoly5eqs} as $z=\sigma+iq_{}$, see \Cref{fig:spectrum5}. 
We find that only the slow mode $s_{4}$ contains points $(q_{}=0,\sigma=\pm a/\lambda)$ in the complex plane with 
\begin{equation}
s_{4}=0\,,
\end{equation}
which defines the steady state. Here, the values of 
$\sigma$ correspond to the decay length given in \cref{teq:coshsigma}. Thus, the shape of the distribution of molecules at steady state is determined by the slow diffusive relaxation mode $s_{4}$. 
And the effective diffusion coefficient and effective degradation rate are  $D_{4}$ and $K_{4}$,
respectively. 
The expansion of mode $s_{4}(z)$ for small $z$ gives a similar relation to \cref{eq:seriesqsmallq},
\begin{equation}
s_{4}\approx K_{4}-\frac{D_{4}}{a^2}z^2\,.
\end{equation}
From \cref{fig:spectrum5}, the zeros of $s_{4}(\sigma+iq_{})=0$ are given at $q_{}=0$, then we find that $\sigma^{2}\approx a^2K_{4}/D_{4}$ for $s_{4}=0$, thus we can relate the decay length of the concentration gradient $\lambda=a/\sigma$ with the effective dynamic parameters of the slow mode $s_{4}$ as
\begin{equation}
\lambda\approx\sqrt{\frac{D_{4}}{K_{4}}}\,.
\end{equation}
This approximation is valid as long as $q_{}$ and $\sigma$ are small, which implies that the decay length must be large.

\bibliographystyle{apsrev4-2}
%

\end{document}